\documentclass[10pt, conference, compsocconf]{IEEEtran}

\usepackage{graphicx}
\usepackage[tight,footnotesize]{subfigure}
\begin{document}
\title{Secure IoT access at scale using blockchains and smart contracts}

\author{\IEEEauthorblockN{Nikos Fotiou, Iakovos Pittaras, Vasilios A. Siris, Spyros Voulgaris, George C. Polyzos}
\IEEEauthorblockA{Mobile Multimedia Laboratory,\\
Department of Informatics School of Information Sciences and Technology\\
Athens University of Economics and Business, Greece\\
\{fotiou,pittaras,vsiris,voulgaris,polyzos\}@aueb.gr}
}

\maketitle

\begin{abstract}
Blockchains and smart contracts are an emerging, promising technology, that has received considerable attention. We use the blockchain technology, and in particular Ethereum, to implement a large-scale event-based  Internet of Things (IoT) control system. We argue that the distributed nature of the ``ledger,'' as well as, Ethereum's capability of parallel execution of replicated ``smart contracts'', provide the sought after automation, generality, flexibility, resilience, and high availability. We design a realistic blockchain-based IoT architecture, using existing technologies while by taking into consideration the characteristics and limitations of IoT devices and applications. Furthermore, we leverage blockchain's immutability and Ethereum's support for custom tokens to build a robust and efficient token-based access control mechanism. Our evaluation shows that our solution is viable and offers significant security and usability advantages. 
\end{abstract}


\IEEEpeerreviewmaketitle

\section{Introduction}
Blockchain technology is expected to revolutionize and ``democratize'' the Internet of Things (IoT)~\cite{ibm2014}, facilitating alternative communication paradigms and enabling novel security mechanisms~\cite{fot2018}. The solutions presented in this paper are a step towards this direction: we take advantage of the  distributed nature of  blockchains to build a large scale IoT control system, and we leverage smart contract based tokens to implement a novel access control mechanism. We argue that existing approaches lack realism and do not take full advantage of the possibilities and capabilities of the blockchain technology. Indeed, related work in this area either neglects the limitations of the IoT devices, or tries to introduce  new, hard to deploy, blockchain technologies, or proposes(unrealistic) modifications to existing blockchain architectures. Similarly, it does not create new solutions using the new features provided by this novel paradigm, instead it tries to merely transfer existing techniques into the new environment. Although, the latter approach may seem to have some value, it turns out that many of the existing solutions do not consider the particularities of the blockchain technology. For example, (public) blockchains cannot be used for storing secret and sensitive information, nevertheless, many proposals use blockchains for storing private user data and business roles and structures.    
  
The work in this paper is concerned with the secure operation of (large) IoT deployments and is based on the observation that many blockchain solutions can be used as event-based systems. With this in mind we design a blockchain-based architecture that allows users to control IoT devices organized in ``groups'' (e.g., turn on the lights of a smart city). Our architecture, which is built using the Ethereum blockchain~\cite{Wood2014}, considers the limitations and capabilities of the IoT devices, as well as the properties of the blockchain technology. Then, we secure this architecture by adding a token-based access control solution, using Ethereum's custom tokens. This approach has some significant advantages compared to existing token-based approaches, with the most important being that it is impossible for a user to transfer his security tokens to another user. The contributions of this paper are the following.
\begin{itemize}
\item We design a blockchain-based IoT architecture based on existing technologies and we define its actors and their interactions
\item We design, implement, and evaluate an event-driven IoT management solution based on Ethereum's smart contracts
\item We leverage Ethereum's support for custom tokens to implement a token-based access control mechanism for our management system
\item We design various extensions to our access control mechanism that achieve common security tasks
\end{itemize}   

The remainder of the paper is structured as follows: In Section II we present some background information of the technologies used in our paper, as well as related work. In Section III we introduce our blockchain-based IoT architecture and in Section IV we present a token-based access control system for this architecture. We evaluate our solution in Section V and we conclude our paper in Section VI.  
\section{Background and related work}
Blockchain systems are distributed-ledger architectures where a set of mutually untrusted nodes can agree on a common view of an indelible, tamper-proof, append-only ledger. In its basic form, a ledger includes a list of transactions among users. Users can send new transactions to the blockchain network and, if these transactions are valid, they are eventually appended to the ledger by a randomly selected specialized node referred to as the \emph{miner}. Advanced forms of ledgers  may also include programs known as \emph{smart contracts}. Smart contracts are associated with some ``state'' also stored in the ledger. Users can interact with a smart contract using transactions and they may modify the contract state. 

A popular blockchain architecture that supports smart contracts is Ethereum~\cite{Wood2014}. From a high level perspective Ethereum smart contracts can be regarded as programming classes and users can interact with the public functions of those classes using transactions. Smart contracts are stored in the ledger and they are identified by an address. Moreover, once they are deployed their code cannot be modified. Contracts are implemented in a low level Turing-complete language and they are executed in a ``virtual machine'' known as the Ethereum Virtual Machine. Smart contracts can receive input only from the ledger, other smart contracts, and the user who invoked them, i.e., smart contracts do not have access to information and resources outside the Ethereum blockchain. Some modifications to a smart contract state can be marked as ``events'' and end-user applications and libraries that monitor the Ethereum blockchain can ``raise alerts'' whenever a specific type of event occurs.

Ethereum users own (at least) one public-private key pair. The private key, which is usually protected in a ``wallet,'' is used for signing transactions. A user may own an Ethereum ``full node'' and interact directly with the blockchain, or he may relay his transactions through another full node that also acts as a ``remote procedure call (RPC)'' server. Each design choice has its trade-offs: maintaining a full node requires continuous network connectivity and some non-negligible storage space for storing the Ethereum blockchain,\footnote{The size of the Ethereum blockchain on 28 Feb. 2019 was reported by https://etherscan.io to be 132GB} whereas relaying transactions through an RPC server entails the risk that the RPC server is offline or it acts maliciously and drops messages.\footnote{An RPC server cannot act on a user's behalf, neither can it replay messages.} 

Early attempts to incorporate blockchain technology into the IoT proposed new blockchain systems. For example, Dorri et al.~\cite{Dor2017} designed a blockchain-based smart home management system. They proposed a custom, blockchain technology, where the home gateways hold the role of the miners. Such solutions are hard to be deployed since they require a ``critical mass.'' Our approach is built on existing technologies and can be used with already available libraries and wallets.

More recent attempts are using blockchain technology and smart contracts to provide security and access control for the IoT.  Hammi et al.~\cite{Ham2018} propose a  blockchain-based IoT communication system. They use an Ethereum smart contact to group IoT devices in ``bubbles'' of trust. Each bubble is managed by a ``master'' which decides which device can join the bubble. In order for a device to join a bubble it must present to the smart contract a ``lightweight certificate'' signed by the bubble master. After joining a bubble a device can communicate with the rest of the bubble members. The communication can take place only through the smart contract, which checks if the sending and receiving devices belong to the same bubble. Novo~\cite{Nov2018} proposes a blockchain-based architecture for managing access to IoT devices. The proposed solution is based on an Ethereum smart contract where ``managers'' can define the IoT resources that another device can access. Gateway nodes, called ``management hubs'', are responsible for handling resource requests by taking into consideration the policies stored in the blockchain.  Zhang et al.~\cite{Zha2019} propose a smart contract based access control system for the  IoT. In their construction the actions a ``subject'' can perform on an ``object,'' as well as the corresponding permissions are recorded in an ``access control contract''. A ``register contract'' is responsible for maintaining a mapping from subject-object identifier pairs to access control contract addresses. An IoT gateway handles resource requests and is responsible for enforcing the access control policies defined in the corresponding access control smart contract. Those solutions follow a similar pattern: they encode in a smart contract the actions a specific user can perform to a particular IoT device/resource. Our solution extends these approaches by considering the token balance of users in the access control policies. In other words, those solutions resemble to an access system which is based on usernames and passwords, whereas our solution resembles a role-based access control system. Furthermore, by leveraging the token handling functionalities of the Ethereum platform our approach enables some novel constructions. 

Recently, Hanada et al.~\cite{Han2018} explored the potential of smart contracts for machine-to-machine (M2M) communication. To this end, they developed and evaluated an IoT application for automated, M2M, gasoline purchases that uses Ethereum smart contracts to perform transactions. Our work is also in this direction. Nevertheless, in addition to merely using smart contracts to provide message transfer and payments, our solution supports group communication and access control.   

\section{A blockchain-based IoT architecture}

\begin{figure*}
\centering
\includegraphics[width=0.5\linewidth]{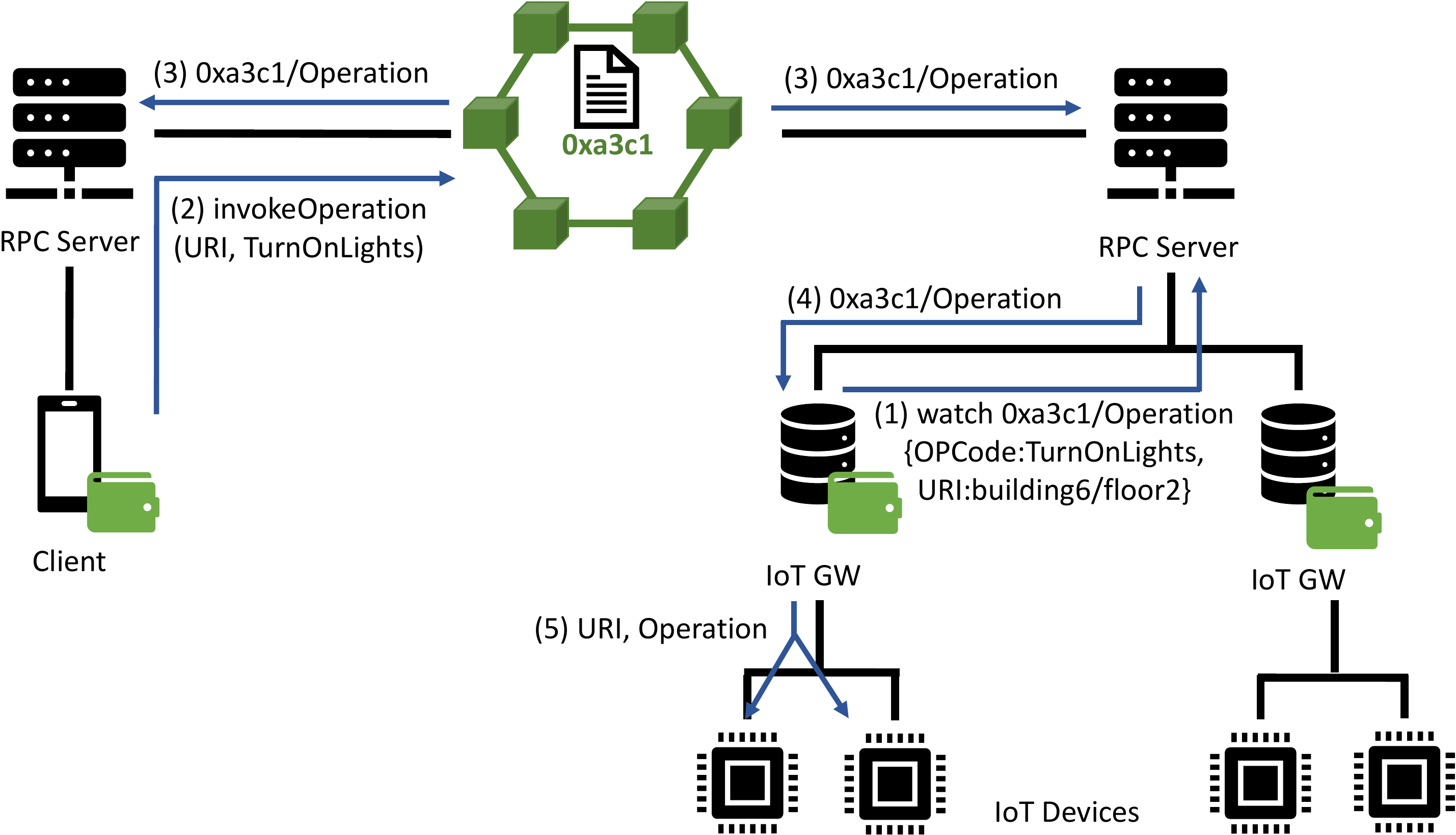}
\caption{Our blockchain-based IoT architecture.}
\label{fig1}
\end{figure*}

We now describe our Ethereum-based IoT architecture (as the typical smart-contract enabled blockchain architecture). Our architecture is composed of the following entities:
\begin{itemize}
\item The blockchain infrastructure
\item A smart contract that generates events
\item Full nodes that also act as RPC servers
\item IoT devices
\item IoT gateways
\item Clients that want to control the IoT devices
\end{itemize}
Clients and IoT gateways are in control of an Ethereum blockchain wallet. A client does not have to interact directly with an IoT gateway (or IoT device), instead all interactions take place through the blockchain. From a high-level perspective our architecture is designed as follows. All device operations are mapped to a function in a smart contract; every time a client invokes a function (properly) the smart contract generates the corresponding blockchain events. These events are received by interested IoT gateways and eventually result in an operation in the appropriate IoT devices.

Clients and IoT gateways can be Ethereum full nodes themselves or they can be connected to the blockchain through another full node acting as an RPC server.  In the following we consider the latter design option. IoT devices on the other hand are connected to IoT gateways. In this section we do not consider any particular governance model; any IoT gateway may ``watch'' for (and act upon) events and any Ethereum user can act as the system client. In the following section we describe an access control mechanism where only authorized users can act as a client.

As IoT devices we consider actuators and we assume that an actuation process can be invoked through an ``operation'', e.g., ``turn on the light''.\footnote{
Of course, sensors can easily be handled and their operations can be thought of as ``provide me your current data. ''} Furthermore, IoT devices are identified by URIs. Following the semantics of CoAP group communication~\cite{rfc7390} we consider that an IoT device may have multiple URIs and a URI may correspond to multiple devices. The semantics of a URI are application specific, for instance they may indicate the physical location of a device, e.g., ``buidling6/floor3/room2''.  An IoT gateway knows the URIs and the supported operations of the devices attached to it (e.g., by using an out-of-band configuration mechanism, or by using a service discovery protocol--such as~\cite{Ams2019}). 

The main component of our system is a smart contract whose address is considered well-known. When invoked, this smart contract generates the appropriate events. An Ethereum event has a name and some attributes. An RPC client may request to watch the events produced by a smart contract by specifying the event name and optionally a filter over (a maximum of three) ``indexable'' attributes. In our architecture we consider a generic event name (i.e., \emph{Operation}) and we specify for each event two attributes: an indexable called \emph{OPCode} that encodes the desired operation and a second one, also indexable, called \emph{URIResource} that corresponds to the URI of the device(s) in which \emph{OPCode} is applied.\footnote{In order to be more precise, since Ethereum does not allow strings to be indexable, the attribute URIResource holds the hash of the URI.}  IoT gateways register to their RPC server to watch the event \emph{Operation} of our smart contract and (optionally)  specify filters on the event's attributes. 

Clients simply interact with the smart contract and invoke the appropriate functions. The main function of our smart contract is called \emph{invokeOperation}. This function accepts two input parameters: an \emph{OPCode} and a \emph{URIResource}, and generates an \emph{Operation} event whose attributes have the same value as the function call parameters. Eventually, this event reaches the IoT gateways that are ``watching'' for it. In return, each IoT gateway invokes the corresponding operation at the IoT devices that are associated with the specified URI.
An overview of our approach is illustrated in Figure 1. In this figure, there is a client, two IoT gateways, and two IoT devices attached to each gateway. One of the gateways starts watching for the \emph{Operation} event of the smart contract located at the address ``0xa3c1'' (step 1). Furthermore, the gateway requests events to be filtered based on their \emph{OPCode} and specifies that it wants to watch only for events in which \emph{OPCode} is ``TurnOnLights''. At some point a client invokes the \emph{invokeOperation} function of the smart contract. It uses as \emph{URIResource} a URI that matches the IoT devices of the aforementioned gateway and as \emph{OPCode} ``TurnOnLights'' (step 2). This transaction results in the creation of an event, which is propagated to all full nodes (step 3). Furthermore, it is transmitted to the IoT gateways that are watching for such events, including our example gateway (step 4). The gateway extracts the \emph{URIResource} of the event and checks if it matches any of the IoT devices attached to it. Since this is the case in our example, the IoT gateway executes the corresponding operation on the appropriate devices (step 5).

\section{Token-based access control using smart contracts}
Many legacy access control mechanisms implement access control using ``tokens'' that indicate the capabilities of a client over a resource. However, token management, security, and semantics interpretation cannot be trivially implemented, especially in the context of the IoT. For this reason, in this section we leverage the capability of the Ethereum blockchain to support custom tokens and we implement an access control mechanism.

Ethereum has specified a ``token standard'' called ERC20~\cite{Vol2015}. This standard defines some functions that a smart contract should implement in order to be treated as a token (i.e., a new type of coin). Many popular Ethereum wallets can handle ERC20-based tokens. The core of our access control mechanism is built using two of these functions, namely \emph{balanceOf} and \emph{transfer}. The first function returns the token balance of a user. The second function can be invoked by a user A in order to transfer some tokens (he owns) to another user B.
 
The smart contract of the architecture defined in the previous section is extended with implementations 
of the functions defined by the ERC20 standard. These extensions can be used for providing access control as follows. Initially a user that owns the smart contract assigns all tokens to himself. We refer to this user as the ``owner''.  The owner then transfers at least one token to each authorized client. As a matter of fact, the number of tokens a client owns can be used as an indication of his role: the more tokens he owns the more privileged his role. The contract owner can protect an operation by specifying the roles (i.e., the balance in custom tokens) of the authorized clients. Therefore, in the simplest case, an operation can be protected simply by having the smart contract function checking if the client that invokes it owns the necessary number of tokens (this check is trivially implemented using the \emph{balanceOf} function). We now discuss some more advanced applications of our approach.

\subsection{Token transfer}
In theory, and based on the ERC20 semantics, any client can transfer some of his tokens to another client using the \emph{transfer} function. Of course, this constitutes a security threat since this way a client authorizes another client--potentially malicious--to perform an operation.  It should be noted here that this is an existing threat in legacy token-based access control systems. Fortunately, ERC20 defines only an ``interface'' and does not dictate any particular implementation choice. Hence, in our contract, a client is allowed to transfer his tokens only to the owner. This transfer is enabled in order to support functionalities such as ``shifts'' where a client is authorized to perform an operation only for a specific time period (that corresponds to his shift) and then transfers through the owner his authorization to the client of the next shift. It should be noted here that off-chain token transfers are impossible. 

\subsection{Clients in probation period}
Another interesting capability of an ERC20 compatible smart contract is that it can modify the token balance of a user at will. In our mechanism we leverage this feature to support clients in probation, trainees, and similar roles. In particular, we allow the owner to define a list of clients whose balance is decreased by one every time they invoke an operation. This way these clients are allowed to perform only a certain number of operations, then the results of these operations are inspected (out of band), and if everything is as expected, the clients regain their tokens back.
 
\subsection{Supervised operations}
Using our mechanism, it is possible to define ``critical'' functions, that require the ``approval'' of a client that holds a more privileged role. In particular, if such a function is invoked by an underprivileged client, instead of producing an \emph{Operation} event, a new type of event is produced called \emph{AuthorizationRequest}. This event is handled by a privileged client, who inspects its fields and acts accordingly, i.e., he may ignore it, or he may invoke the same function again so that the \emph{Operation} event is generated. 

\subsection{Two-steps access control}
Since the Ethereum ledger is distributed (which is a key property of blockchain-based systems) any full node (or RPC client) can learn the token balance of a user without interacting with the corresponding smart contract. This property enables the definition of additional (possibly finer grained) access control policies at the IoT gateways. This means that even if a client is authorized by the smart contract, eventually his operation may be rejected by some/all IoT gateways. The access control policies defined at the IoT gateways may take into consideration, in addition to the role of the client, other auxiliary information provided by the ``real'' world, such as time, location information, other IoT measurements, etc. Notice that smart contracts do not have access to such information.

\subsection{Panic button}
Our access control smart contract defines a function that can be invoked only by the owner and it resets the token balances of all users, returning in essence all tokens back to the owner. This function can be used in case of emergency, e.g., in case of a security breach. Additionally, when invoking this function, the owner can specify the public key of a user, resetting this way the balance of that particular user. Using this approach client revocation can be trivially implemented. Since all transactions are recorded in the blockchain, it is painless to restore user balances to their value prior the ''panic button'' invocation. Moreover, the clients whose tokens are revoked have no control over this process, hence revocation is instantaneous and effective. 

\section{Evaluation and Discussion}
\subsection{Performance and cost evaluation}
We have implemented and tested our proposed solution in a private Ethereum network, as well as in the Rinkeby and Ropsten Ethereum testnets. As an IoT gateway we have used Mozilla's Thing Gateway\footnote{https://iot.mozilla.org/} that implements the Web of Things standard.\footnote{https://www.w3.org/WoT/} We implemented clients as JavaScript web applications using web3.js Ethereum JavaScript API\footnote{https://web3js.readthedocs.io} and the Metamask Firefox extension.\footnote{https://metamask.io/}. 

The invocation of an Ethereum smart contract function creates some computational overhead measured in ``gas'' units: the amount of gas ``consumed'' by a function depends on the operation's complexity. Each user declares the price he is willing to pay per gas unit: the bigger the amount, the faster the operation will be executed. The fastest an operation can be executed is $\sim$14sec. which is the time required by the Ethereum network to generate a new block. Hence, users compete each other since they wish to execute their operation fast but they do not want to get charged a lot. Currently, the average price of a unit of gas is\footnote{As measured by https://ethgasstation.info on 20 Mar. 2019} \$$0.004 \times 10^{-4}$. Our construction uses Ethereum's events and it is built using a ``mapping'' type, i.e., a hashtable-like data structure that maps ``keys'' to ``values''. Our events have two fields, namely \emph{OPCode} of type \emph{byte} and \emph{URIResource} of type \emph{bytes32} (i.e., a byte array of size 32). Furthermore, our mapping maps keys of type \emph{address} to values of type \emph{int} and it is used for maintain client's balance.  The primitive operations required by our smart contract are map \emph{search}, \emph{creation} of a new map entry, and \emph{modification} of the value of a map entry. Table 1 shows the cost of these operations in terms of gas consumption. In addition to these costs, each transaction has an overhead of 21000 gas. 

\begin{table}[h]
\centering
\begin{tabular}{ | l | c |}
\hline
\textbf{Operation} & \textbf{Cost measured in gas} \\ \hline
Send \emph{Operation} event     & 2560                             \\
Search map     & 1033                              \\
Map entry creation   & 45938                              \\
Map entry modification    & 6110                              \\ \hline
\end{tabular}
\caption{Cost of our construction building blocks}
\end{table}

\subsection{Qualitative and security properties}
Our construction leverages the inherent properties of the blockchain technology. By design, blockchain solutions offer reliability and robustness, since the ``ledger'' is replicated in multiple locations and there is no single point of failure. Furthermore, blockchain communication protocols and APIs include  message integrity protection, as well as resilience against replay attacks. Smart contract execution is deterministic and cannot be affected by malicious entities. Similarly, smart contracts cannot be modified, not even by their owners. As already discussed, invoking a smart contract function has some monetary cost; this could be an effective defense against Denial of Service attacks.

As far as our token-based access control is concerned, it can be observed that it has some intriguing security properties. Firstly, tokens can only be used by their owners, and token owners cannot transfer them to other users. Even if the blockchain keys of a user are compromised, our construction prevents token transfer (of course the stolen keys can be used for issuing transactions on behalf of the victim users). This is a significant advantage compared to traditional token-based access control mechanisms where, not only the corresponding tokens have to be secured, but also a token recipient should be able to verify the binding between the token and the user who sent it (i.e., additional mechanisms for detecting stolen tokens should be in place). In other words, the responsibility (and security) of binding of tokens to token owners is performed by the blockchain and it is not the responsibility of each user (which opens security issues). Furthermore, and as already discussed, blockchains are an indelible, append-only, and tamper-proof logs, hence, in case of a security incident or in case of a dispute they can provide undeniable auditing information. Moreover, our construction offers secure and effective revocation. Ethereum's mechanisms guarantee that only an owner can revoke tokens (providing of course that the owner's private key is secured), as well as, that a token revocation has immediate affect. Finally, since our construction is based on an established Ethereum standard, libraries and wallets that support it, can be used for implementing client applications.     

Ethereum is composed of a P2P network where all valid blocks are broadcast to all nodes. In reality, events are special fields encoded in those blocks, hence the number of nodes watching for events does not have any impact on the number of transmitted messages. In other words, if we take Ethereum infrastructure for granted (or any other similar architecture) it is costless to build a group communication application on top of that. Moreover, due to this property, the network location of the IoT gateways does not have to be well known, neither have gateways to be reachable through the Internet. 

\subsection{Discussion}
Despite the advantages of the blockchain technology, it comes with some costs. As already discussed Ethereum (and most blockchain systems) involve some monetary cost, as well as some transaction delay. Unfortunately, and since Ethereum is still an experimental technology, the monetary cost of transactions fluctuates greatly. Furthermore, Ethereum operates on the premise that at least half of the network nodes are honest; having an attacker controlling more than $50\%$ of the nodes in an unlikely but not impossible threat. Finally, Ethereum's ledger is public and anybody can inspect it. This property constitutes a privacy threat since it is possible for a thrid party to deduce information such as, who perform which operation and when, the ``roles'' of the users, the introduction of new authorized users, etc. All these shortcomings can possibly  be  addressed using a ``permissioned'' private blockchain, such as Hyperledger Fabric.\footnote{https://www.hyperledger.org/projects/fabric}

In the construction presented in this paper clients interact with the IoT devices only through the blockchain. Of course cases where a client interacts directly with an IoT gateway can be considered. This direct interaction has some advantages, including zero transaction fees and faster response times. Moreover, and since the Ethereum ledger is replicated in all nodes, a gateway can still perform token-based access control. On the other hand, in this case, the gateway should be able to verify the identity of the client (i.e., his blockchain public key).

\section{Conclusion}
In this paper we designed, developed, and evaluated an IoT access architecture based on smart contracts and blockchains. Our solution leverages the distributed nature of the blockchain technology to build an event-based system for managing IoT devices connected to Web of Things gateways. Furthermore, we enhanced our architecture with an access control solution based on custom blockchain tokens. Our access control solution has some intriguing properties and presents some important advantages compared to traditional token-based access control systems. Finally, our Ethereum-based implementation shows that our solution is feasible and with low overhead. 

Blockchain and smart contracts are an exiting, evolving technology, with endless possibilities. Hence, our system can be extended in numerous ways. For instance, our system can be extended to support an IoT-based sharing economy, or even auctions over IoT access tokens (e.g., a frivolous but prosperous use case could be an auction for the token that can light the Christmas tree of a city). Similarly, the blockchain can be used for tracking user reputation or even ``score'' in a gameficated application. Finally, our architecture can be extended to support blockchain-based decentralized identifiers, a new technology under standardization with exciting security and privacy properties. 

\section*{Acknowledgments}
This research was supported by the EU funded Horizon 2020 project SOFIE (Secure Open Federation for Internet Everywhere), under grant agreement No. 779984. Iakovos Pittaras was supported by the Athens University of Economics and Business Research Center.     

\bibliographystyle{IEEEtran}

\bibliography{IEEEabrv,iot-sos}
\end{document}